\documentclass[a4paper]{article}
\usepackage[a4paper,includemp=false,body={16cm,24.7cm},vcentering]{geometry}
\usepackage[latin1]{inputenc}
\usepackage[T1]{fontenc}
\usepackage{mathptmx} % police Times texte et maths
\usepackage[bf,center]{caption}
\usepackage{graphicx} % si figures

\usepackage{url,amsmath,amssymb,amsthm}

\makeatletter
\renewcommand{\section}{\@startsection{section}{1}{0mm}{30pt}{12pt}{\normalfont\normalsize\bfseries}}

\renewcommand{\subsection}{\@startsection{subsection}{2}{0mm}{18pt}{12pt}{\normalfont\normalsize\itshape}}
\makeatother

\newcommand{\Title}[1]{\begin{center}{\bfseries\fontsize{12pt}{12pt}\selectfont#1}\end{center}}
\newcommand{\Author}[2]{\begin{center}{\fontsize{12pt}{12pt}\selectfont#1}\\{\it #2~}\end{center}}
\newcommand{\Introduction}{\section*{Introduction}}
\newcommand{\Conclusion}{\section*{Conclusion}}

\begin{document}

\Title{Astrometry and Exoplanet Characterization: Gaia and Its Pandora's Box}
\vskip 0.5cm  
\Author{Sozzetti, A.$^1$ }{1. INAF - Osservatorio Astronomico di Torino \\ Via Osservatorio 20, I-10025 Pino Torinese, Italy \\ sozzetti@oato.inaf.it}
 
\Introduction

In its all-sky survey, Gaia will monitor astrometrically hundreds of thousands of main-sequence stars within $\approx200$ pc, 
looking for the presence of giant planetary companions within a few AUs from their host stars. Indeed, Gaia observations will 
have great impact is the astrophysics of planetary systems (e.g., Casertano et al. 2008), in particular when seen as a complement 
to other techniques for planet detection and characterization (e.g., Sozzetti 2011). 
In this paper, I briefly address some of the relevant technical issues associated with the precise and accurate determination of astrometric 
orbits of planetary systems using Gaia data. I then highlight some of the important synergies between Gaia high-precision 
astrometry and other ongoing and planned, indirect and direct planet-finding and characterization programs, both from the ground and in space, 
and over a broad range of wavelengths, providing preliminary results related to one specific example of such synergies.

\section{Exoplanet Orbits with Gaia Astrometry: The Challenge}

We are about to enter the age of micro-arcsecond ($\mu$as) astrometry with Gaia. In the matter of astrometric 
detection of planetary-mass companions around nearby stars, the improvement of over two orders of magnitude in single-measurement precision 
with respect to present-day Hipparcos astrometry is absolutely mandatory to finally put an end to the long list of `blunders' that this 
technique has delivered (for a review, see Sozzetti 2010). However, the improved accuracy of the Gaia measurements will not resolve 
by itself the technical problems associated with the modeling of the astrometric signatures of planetary systems, which will have to 
be carefully dealt with. 

\subsection{Orbital Fits of Planetary Systems}

The problem of the correct determination of the astrometric orbits of planetary systems using Gaia data 
(highly non-linear orbital fitting procedures, with a large number of model parameters) will present many difficulties. 
For example, it will be necessary to assess the relative robustness and reliability of different procedures for orbital fits. 
Consistency checks between different solution algorithms will be mandatory as a way of learning the lessons of 
radial-velocity surveys, that are showing us, particularly in the case of multiple-planet systems, how 
disagreement on orbital solution details, and sometime number of planets!, based on the same datasets is not that uncommon 
(e.g., Forveille et al. 2011, and references therein; Hatzes et al. 2011, and references therein). 
A detailed understanding of the statistical properties of the uncertainties
associated with the model parameters will have to be developed, based on the relative merit of different 
metrics tailored to this task, such as covariance matrices, $\chi^2$ surface mapping, and bootstrapping procedures. 
For multiple systems, a trade-off will have to be 
found between accuracy in the determination of the mutual inclination angles between
pairs of planetary orbits, single-measurement precision and redundancy in the number
of observations with respect to the number of estimated model parameters. It will constitute a challenge to 
correctly identify signals (and the associated) with amplitude close to the measurement
uncertainties, particularly in the presence of larger signals induced by other companions and/or 
sources of astrophysical noise of comparable magnitude. Finally, in cases of
multiple-component systems where dynamical interactions are important (a situation ex-
perienced already by radial-velocity surveys), fully dynamical (Newtonian) fits involving
an n-body code might have to be used to properly model the Gaia astrometric data and
to ensure the short- and long-term stability of the solution (see Sozzetti 2005).

\subsection{Exoplanets Treatment in the Gaia Data Processing Pipeline}	

All the above issues could have a significant impact on Gaia's capability to detect
and characterize planetary systems. For these reasons, within the pipeline of Coordination Unit 4 (object processing) 
of the Gaia Data Processing and Analysis Consortium (DPAC), in charge of the scientific processing of 
the Gaia data and production of the final Gaia catalogue to be released sometime in 2021, a Development Unit (DU437) 
has been specifically devoted to the modelling of the astrometric signals produced by planetary
systems. The DU is composed of several tasks, which implement multiple robust procedures for (single and multiple) 
astrometric orbit fitting (such as Markov Chain Monte Carlo and genetic algorithms) and the determination of the 
degree of dynamical stability of multiple-component systems. I provide here a quick-look view of the software and its 
status. 

\subsection{Fitting Algorithms: Highlights}

\begin{figure}
\centering
\includegraphics[width=0.55\textwidth]{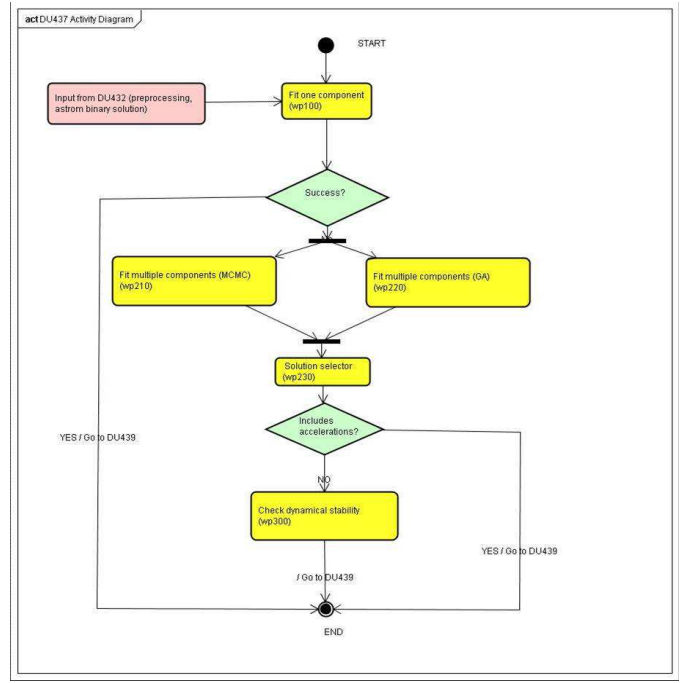}
\caption{DU 437 Activity Diagram.} 
\label{fig1}
\end{figure}

Figure~\ref{fig1} shows the activity diagram of DU437. The main feature of this software module consists of the use 
of two different solution algorithms for fitting astrometric orbits of planetary systems to Gaia data. 

The first algorithm is based on a hybrid Markov Chain Monte Carlo (MCMC) approach. The global non-linear least-square problem is partially 
linearized in then paramters $a_j$, $\omega_j$, $\Omega_j$, and $i_j$ (for $j=1,\dots,n$, where $n$ is the number of planets) 
using the Thiele-Innes elements representation (Green 1985). An iterative period search provides seeds 
for initializing the MCMC procedure (a parallel-tempering algorithm upgrade is in the works). An MCMC chain is then 
run on the non linear parameters ($e_j$, $P_j$, $\tau_j$), drawing from Gaussian independent distributions. 
Each step of the chain drives a linear LSQ solver (SVD). Multiple MCMC chains are run (one is used for inference), and 
convergence is checked via the Gelman-Rubin test. Finally, the set of parameters with the highest likelihood is the seed 
for a local non-linear minimization in the least-squares sense using the Levemberg-Marquardt algorithm. 

The second otbital solution module implements an approach based on a genetic algorithm (GA). The first step consists of 
the generation of a population of $7*n$ chromosomes. The fitness of each chromosome in the population is evaluated, and 
the selection of `parent' chromosome pairs is made from the population according to their fitness. The next step includes 
the iterative modeling using a set of operators such as mutation, cross-over, Levemberg-Marquardt, and period scan. 
Finally, acceptance of new offsprings in the population is realized, and further iterations of the algorithm using the new population 
are carried out. 

The last important element of the DU437 software module is constituted by its orbital stability analysis component. 
At present, a simple Hill stability criterion is applied (Marchal \& Bozis 1982). For a two-planet configuration, 
the system will be considered Hill-stable if the following inequality is satisfied:

\begin{equation}
-\frac{2M}{G^2M^3_\star} c^2 h > 1+3^{4/3}\frac{m_1 m_2}{m_3^{2/3} (m_1+m_2)^{4/3}},
\end{equation}

where $M$ is the total mass of the system, $m_1$ and $m_2$ are the planet masses (the subscript 1 refers to the inner planet), 
$m_3$ is the mass of the star, $G$ is the gravitational constant, $M_\star = (m_1 m_2+m_1 m_3+m_2 m_3)$, $c$ is the total angular momentum of the system, and h is the energy
This is sufficient to double-check dynamically pathologic (while potentially correct in a $\chi^2$ sense!) solution sets 
(e.g., orbit-crossing due to $e\sim1.0$) obtained by the MCMC and GA algorithms. Limited use of an N-body integrator (e.g., Burlisch-St\"oer) is 
being tested at the time of writing. 

\subsection{A Readiness Test}

\begin{figure}
\centering
$\begin{array}{cc}
\includegraphics[width=0.40\textwidth]{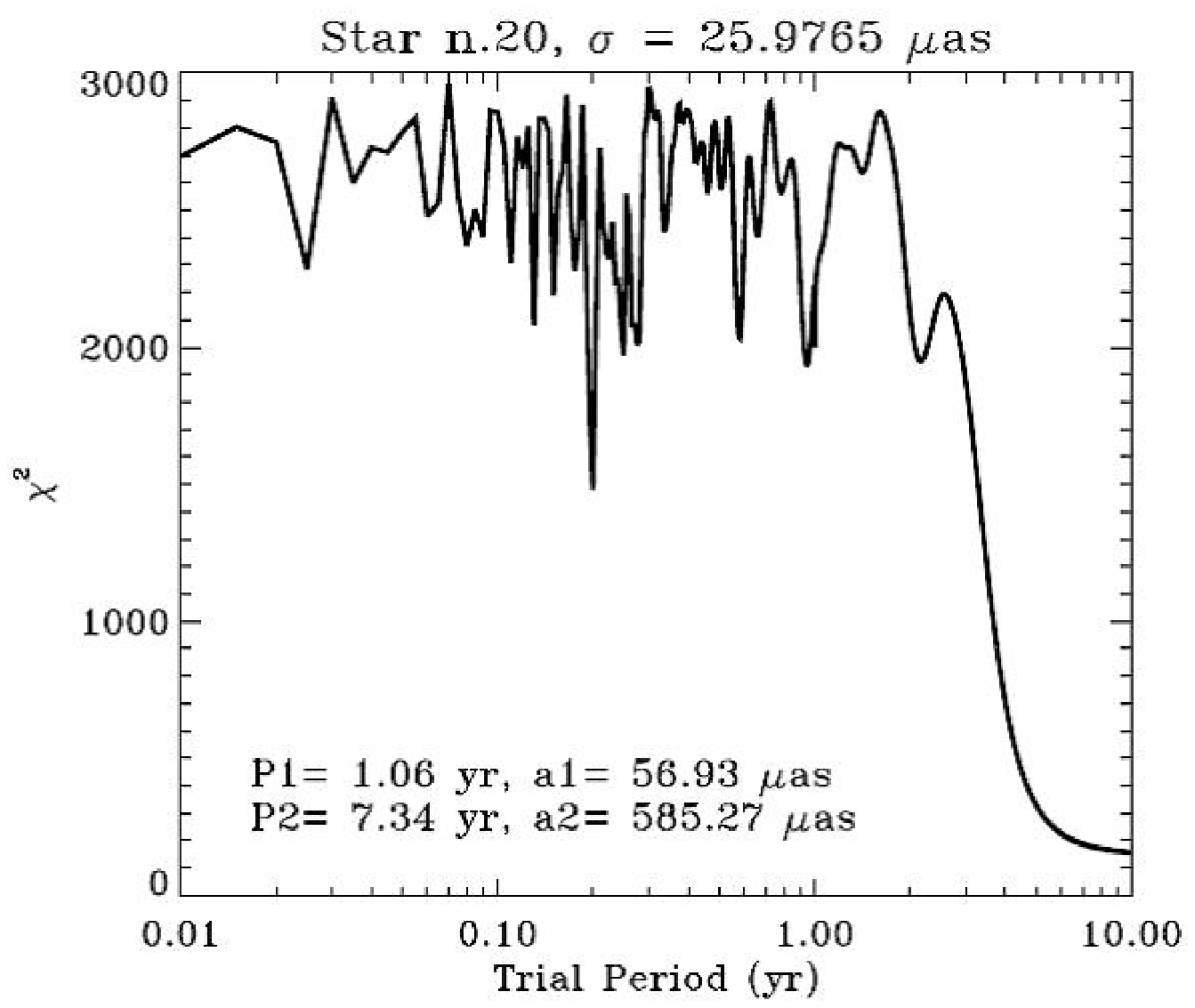} & 
\includegraphics[width=0.55\textwidth]{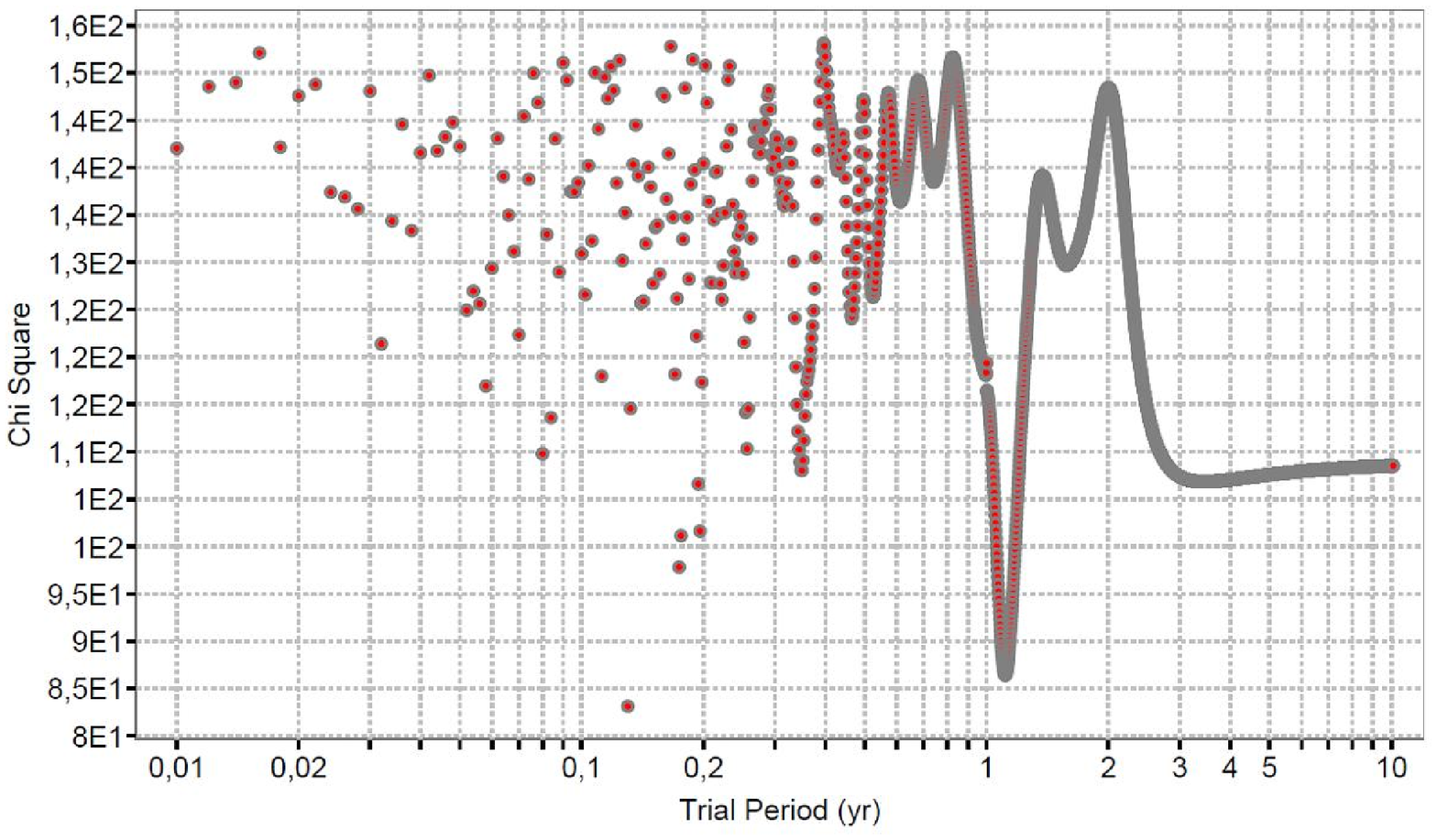} \\
\end{array} $
 \caption{Left: First-pass periodogram for a star with a long-period planet. Right: The periodogram after the removal of 
 the signature of the outer planet.}
\label{fig2}
\end{figure}

\begin{figure}[t]
\centering
$\begin{array}{cc}
\includegraphics[width=0.50\textwidth]{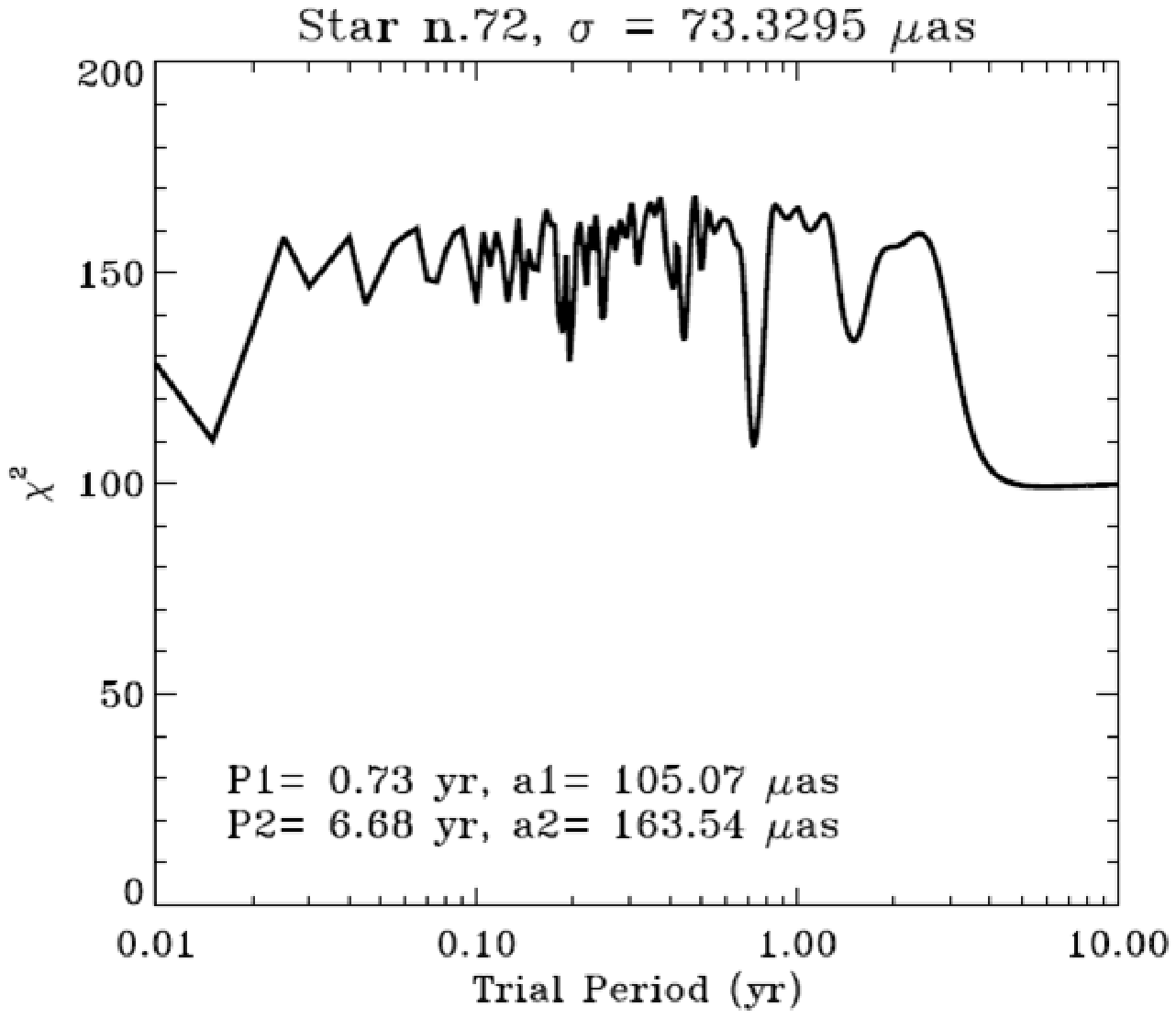} & 
\includegraphics[width=0.40\textwidth]{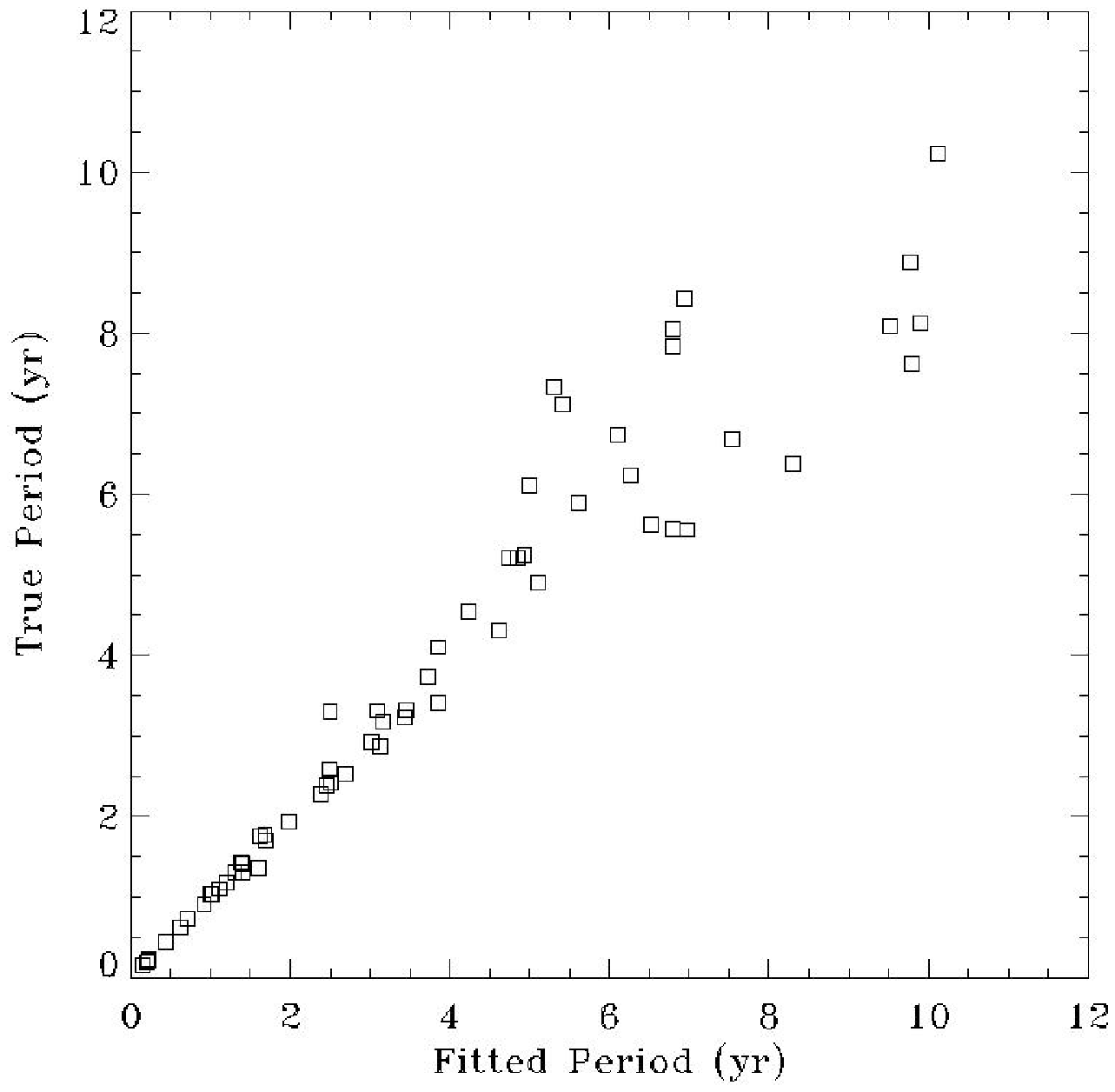} \\
\end{array} $
 \caption{Left: First-pass periodogram for a star with two low-S/N planets. Right: Fitted vs. true values of the orbital period for 
 detectable two-planet systems.}
\label{fig3}
\end{figure}

I show in Figure~\ref{fig2} and Figure~\ref{fig3} some results from the application of the MCMC 
algorithm on a simulated Gaia data for a stellar sample of 100 nearby ($d<100$ pc) solar-mass stars with Gaia magnitude $G<10$ mag and 
orbited by two planets with periods up to twice the nominal 5-yr missione duration. Given the present-day Gaia astrometric error model, 
inclusive of the effects of a gate scheme to avoid saturation at the bright end ($G<13$ mag), in this sample $\approx 1/3$ of 
the systems were not detectable in Gaia astrometry, $\approx1/3$ had one detectable planet, and $\approx1/3$ had two detectable companions. 
For example, in a two-planet system with a long-period planet identified in the initial periodogram (Figure~\ref{fig2}, left panel), the correct 
removal of the outer planet's orbit unveils in the periodogram the clear signal of the second, inner planet (Figure~\ref{fig2}, right panel). 
Similarly, in the case of a system with two planets inducing low-S/N signatures of comparable magnitude, the periodicity analysis correctly 
identifies the periods of both components already in the first pass (Figure~\ref{fig3}, left panel). The resulting quality of the fitted periods 
(shown in Figure~\ref{fig3}, right panel) in detectable two-planet systems is in good agreement with earlier results (Casertano et al. 2008). 

\section{The Gaia - Exoplanets Synergy Potential}

Gaia's main contribution to exoplanet science will be its unbiased census of planetary 
systems orbiting hundreds of thousands nearby ($d < 200$ pc), relatively bright
($G<13$) stars across all spectral types, screened with constant astrometric sensitivity. 
As a result, the actual impact of Gaia measurements in exoplanets science
is broad, and rather structured. The Gaia data have the potential to: a) significantly 
refine our understanding of the statistical properties of extrasolar planets;
b) help crucially test theoretical models of gas giant planet formation and migra-
tion; c) achieve key improvements in our comprehension of important aspects of
the formation and dynamical evolution of multiple-planet systems; d) aid in the
understanding of direct detections of giant extrasolar planets; e) provide important 
supplementary data for the optimization of the target selection for future
observatories aiming at the direct detection and spectral characterization 
of habitable terrestrial planets. For a review, see Sozzetti (2010). 

The broad range of applications to exoplanets science is such that Gaia data can
be seen as an ideal complement to (and in synergy with) many ongoing and future
observing programs devoted to the indirect and direct detection and characterization 
of planetary systems, both from the ground and in space, and across a broad range of wavelengths. 
Gaia will contribute critically, for example, to the definition of input catalogues for proposed quasi-all-sky 
photometric transit surveys (PLATO, TESS); it will inform ground-based direct imaging programs 
(e.g., SPHERE/VLT, EPICS/E-ELT) and spectroscopic characterization projects (e.g., EChO, FINESSE)
about the epoch and location of maximum brightness of (primarily non transiting) 
exoplanets, in order to estimate their optimal visibility, and will help in the
modeling and interpretation of giant planets' phase functions and light curves. 
Another critical aspect will concern the large effort in terms of ground-based
follow-up activities to improve the characterization of astrometrically detected
systems (and possibly those found transiting by Gaia photometry). For example, high-precision radial-velocity 
campaigns (both at visible and infrared wavelengths) will be a necessary
complement, with the three-folded aim of improving the phase sampling of the
astrometric orbits found by Gaia, extending the time baseline of the observations
(to put stringent constraints on or actually characterize long-period companions),
and search for additional, low-mass and/or short-period components which might
have been missed by Gaia due to lack of sensitivity.

\subsection{A Synergetic Example: The Gaia survey of Nearby M Dwarfs}

Cool, nearby M dwarfs within a few tens of parsecs from the Sun are becoming the focus of dedicated 
experiments in the realm of exoplanets astrophysics. This is due to the shift in theoretical paradigms 
in light of new observations, and to the improved understanding of the observational opportunities for 
planet detection and characterization provided by this sample. Gaia, in its all-sky survey, will deliver 
precision astrometry for a magnitude-limited ($G=20$) sample of M dwarfs, providing an inventory of cool 
nearby stars with a much higher degree of completeness (particularly for late sub-types) with respect 
to currently available catalogs. 

I present here preliminary findings of a simulation experiment aimed at gauging the Gaia potential for 
precision astrometry of exoplanets orbiting a sample of known, nearby dM stars (L\'epine 2005). 
Gaia sensitivity thresholds are expressed as a function of system parameters and in view of the latest mission profile, 
including the most up-to-date astrometric error model. The simulations also provide insight on the capability of high-precision astrometry to 
reconstruct the underlying orbital elements and mass distributions of the generated companions. 
These results will help in evaluating the complete expected Gaia planet population around late-type stars. 

The synergy between the Gaia data on nearby M dwarfs and other ground-based and space-borne programs for planet detection and characterization 
is also investigated, with a particular focus on: a) the potential for Gaia to precisely determine the orbital inclination, 
which might indicate the existence of transiting long-period planets; b) the ability of Gaia to carefully predict 
the ephemerides of (transiting and non-transiting) planets around M stars; and c) its potential to help in the precise 
determination of the emergent flux, for systematic spectroscopic characterization of their atmospheres with dedicated 
observatories in space, such as EChO.

\paragraph{Simulation Scheme}

\begin{figure}[t]
\centering
$\begin{array}{cc}
\includegraphics[width=0.45\textwidth]{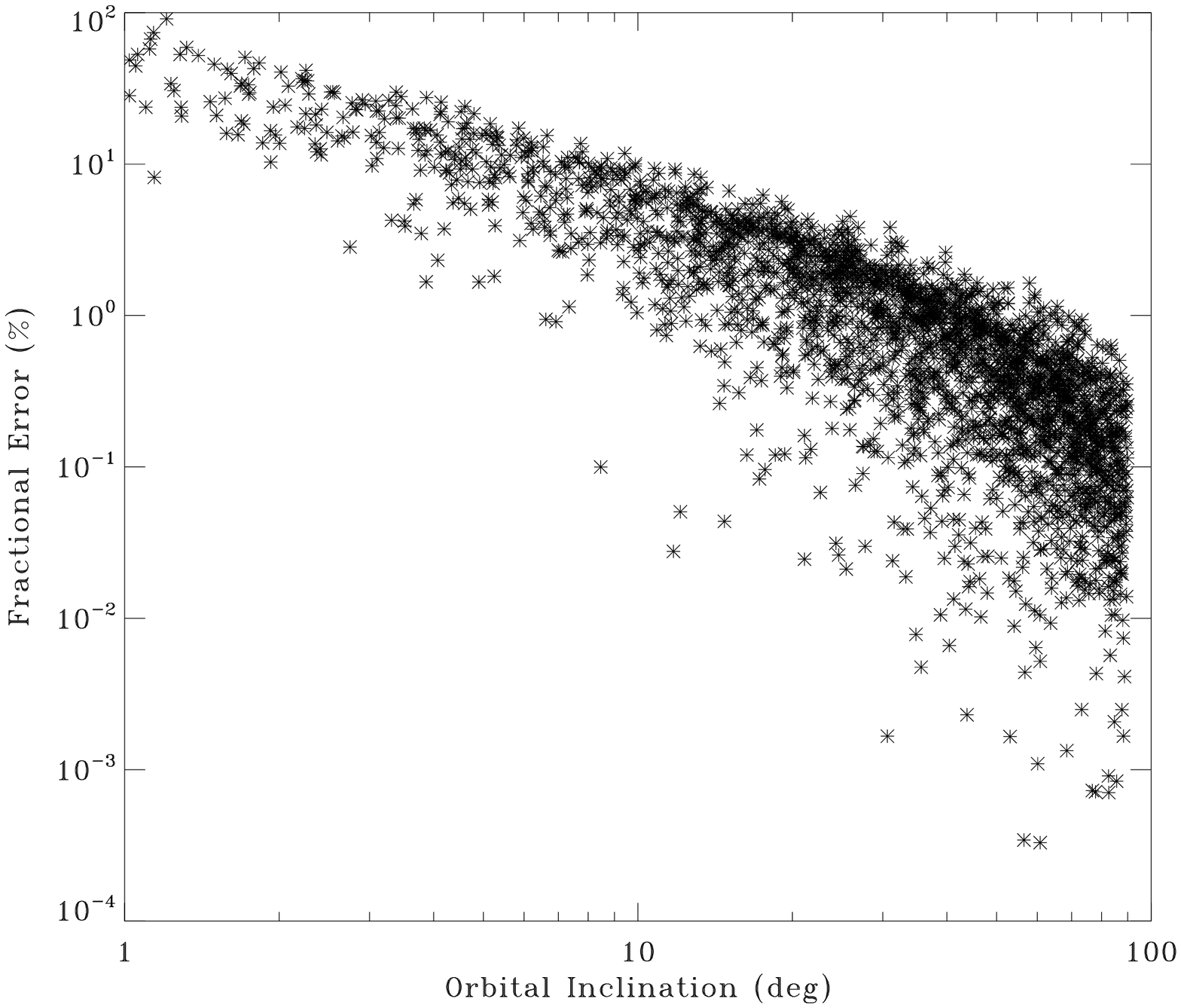} & 
\includegraphics[width=0.45\textwidth]{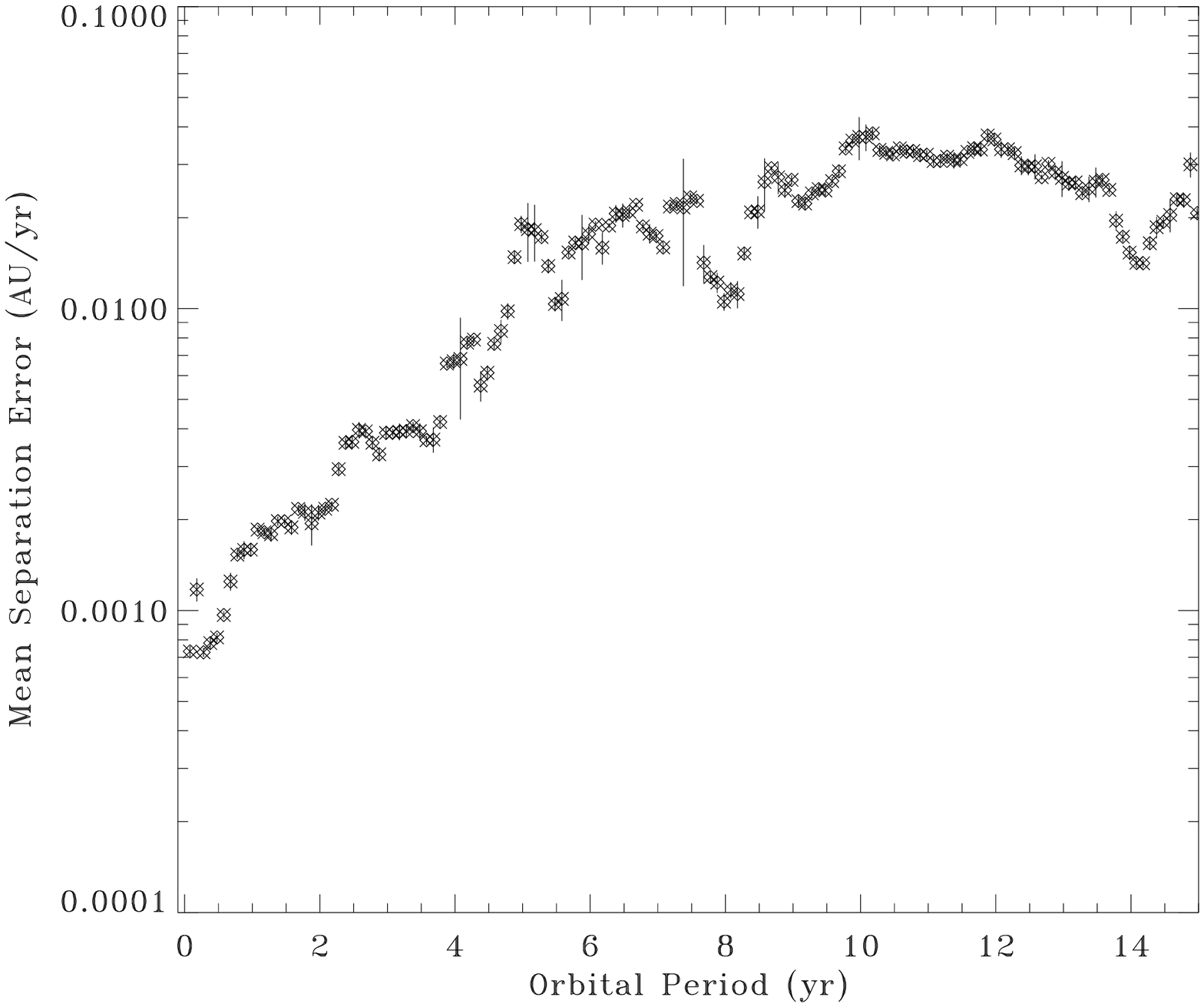} \\
\end{array} $
 \caption{Left: Fractional error on the inclination angle $i$ as a function of $i$ itself. Right: Degradation rate in 
 the prediction for the value of orbital separation of the planet as a function of its orbital period.}
\label{fig4}
\end{figure}

The basis is constituted by the Casertano et al. (2008) simulation setup. This has been updated by including the lates 
Gaia scanning law, for a nominal $T=5$ yr mission duration. The most up-to-date error model as a function of Gaia G-band mag 
has been utilized, with the exclusion of the presently envisioned gate scheme (affecting only some 20\% of bright ($G<12$) M dwarfs). 
Single-measurement errors are typically $\sigma_m\sim100$ $\mu$as. The actual list of targets encompasses 3150 M dwarfs ($0.09-0.6$ $M_\odot$) 
within 33 pc from the Sun from the LSPM-North Catalog (L\'pine 2005), with average $G\sim14.0$ mag. One planet was generated around each star, 
with mass $M_p = 1 M_J$, orbital period $P < 3T$, and moderate eccentricities ($e<0.6$). All other orbital elements were uniformly distributed 
within their respective ranges. 

\paragraph{Preliminary Results}

\begin{figure}[t]
\centering
$\begin{array}{cc}
\includegraphics[width=0.45\textwidth]{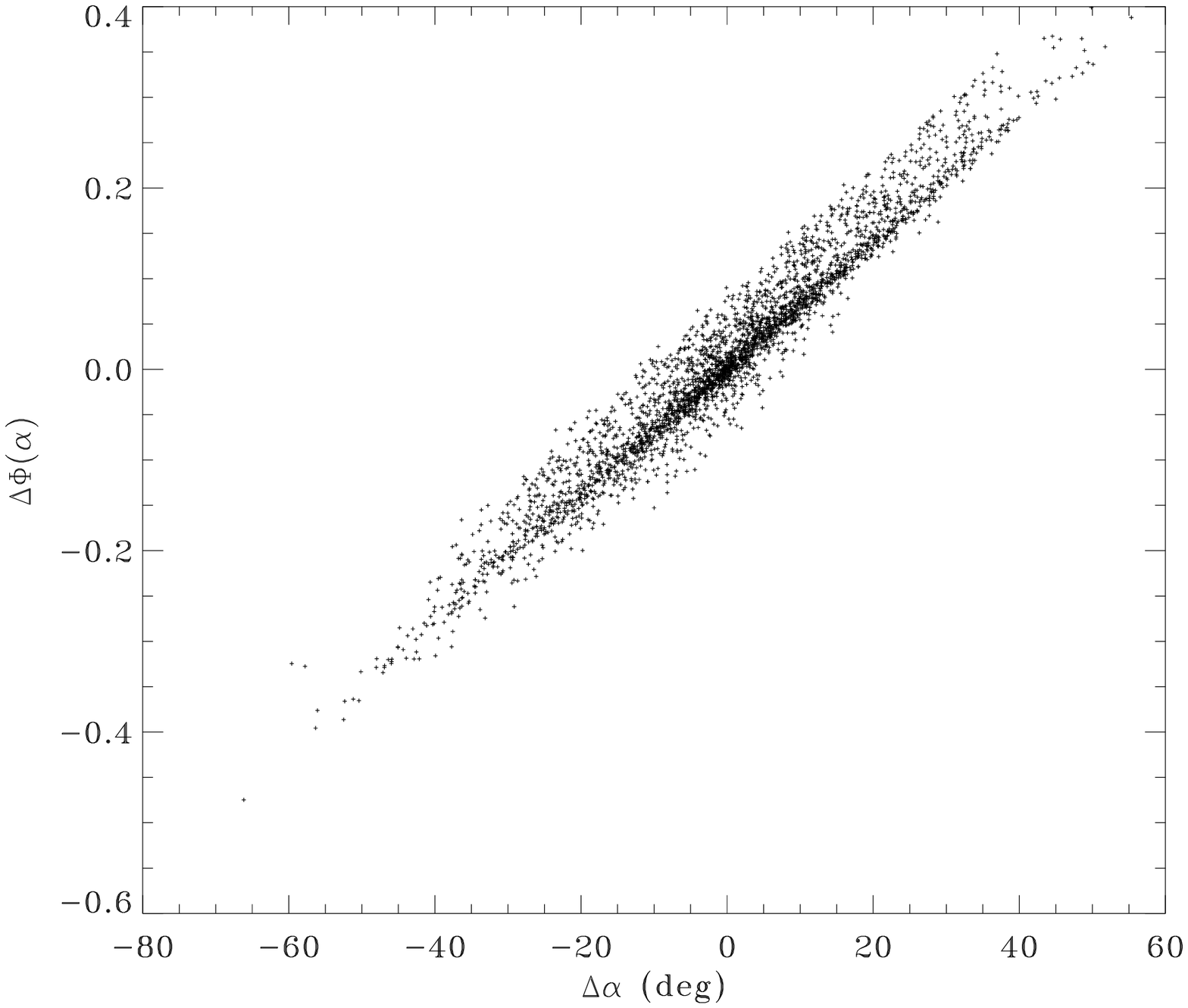} & 
\includegraphics[width=0.45\textwidth]{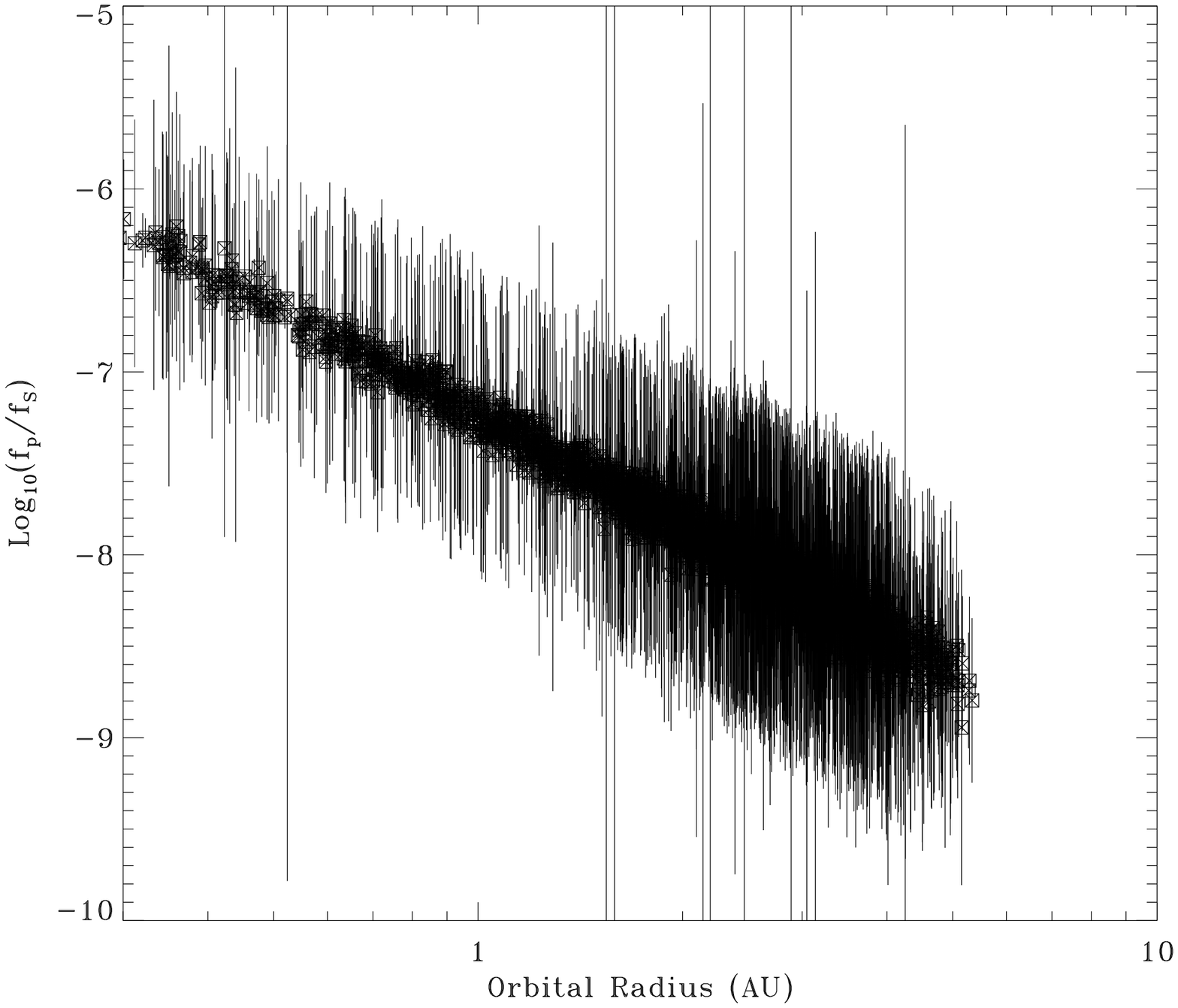} \\
\end{array} $
 \caption{Left: Error on the planetary phase function (assuming a Lambert sphere) as a function of uncertainty on the reconstructed planetary phase. 
 Right: Predicted planetary emergent flux (and uncertainty) as a function of orbital separation from the M dwarf primary.}
\label{fig5}
\end{figure}

The main findings in this experiment can be summarized as follows: 1) For detected giant planets with periods in the range $0.2-5$ yr (i.e., with 
accurately determined masses and orbits), inclination angles will be determined with enough precision (<1\%, see Figure~\ref{fig4}, left panel) 
so that it will be possible to identify long-period planets which are likely to transit; 2) for well-sampled orbits ($P<T$), the uncertainties on planetary ephemerides, 
separation $\varrho$ and position angle $\vartheta$, will degrade at typical rates of $\Delta\varrho < 0.01$ AU/yr (see Figure~\ref{fig4}, right panel) 
and $\Delta\vartheta < 1$ deg/yr, respectively. These are over an order of magnitude smaller than the degradation levels attained by present-day ephemerides 
predictions based on milli-arcsec precision HST/FGS astrometry (Benedict et al. 2006); 3) Planetary phases will be measured with typical 
uncertainties $\Delta\alpha$ of a few degrees, resulting (assuming a simple purely scattering atmosphere) in average errors on the phase function 
$\Delta\Phi(\alpha)\approx0.1$ (Figure~\ref{fig5}, left panel), and expected uncertainties in the determination of the emergent flux of wide-separation 
($a>0.3$ AU) giant planets of $\sim15\%$ (Figure~\ref{fig5}, right panel). 

All the above conclusions help to quantify the actual relevance of the Gaia observations of the large sample of nearby M dwarfs for in a synergetic effort 
to optimize the planning and interpretation of follow-up/characterization measurements of the discovered systems by means of transit survey programs, and 
upcoming and planned ground-based as well as space-borne observatories for direct imaging (e.g., SPHERE, EPICS) and simultaneous multi-wavelength spectroscopy (e.g., EChO).

\Conclusion

The largest compilation of high-accuracy astrometric orbits of giant planets, unbiased across all spectral types up to $d<200$ pc, 
will allow Gaia to crucially contribute to several aspects of planetary systems astrophysics (formation theories, dynamical evolution), 
in combination with present-day and future extrasolar planet search programs. 

In this paper, I have first discussed some of the relevant issue related to the robust assessment of astrometric orbits of planetary systems detected by Gaia, 
with a focus on readiness tests of the DU437 software module in charge of providing (single and multiple) orbital solutions for 
exoplanets within the context of the Gaia data processing pipeline. Then, I have presented preliminary results on an investigation of the synergy 
between Gaia astrometry on nearby M dwarfs and other ground-based and space-borne programs for exoplanet characterization, focusing on: 
a) the potential for Gaia to precisely determine the orbital inclination, which might indicate the existence of transiting long-period planets; 
b) the ability of Gaia to carefully predict the ephemerides of (transiting and non-transiting) planets around M stars; and c) its potential to 
help in the precise determination of the emergent flux, for systematic spectroscopic characterization of their atmospheres with dedicated 
observatories in space. 

\section*{Acknowledgements}

The author wishes to thank M.G. Lattanzi, R. Morbidelli, G. Micela, G. Tinetti, P. Giacobbe and the colleagues of 
(DPAC CU4) DU437 D. S\'egransan, D. Sosnowska, and N. Rambaux for their help, discussions, and inputs.


\begin{thebibliography}{99}

\bibitem{Benedict06} Benedict G.F., et al. 2006. The Extrasolar Planet $\varepsilon$ Eridani b: Orbit and Mass, AJ, 132, 2206. 

\bibitem{Casertano08} Casertano S., Lattanzi M.G., Sozzetti A., et al. 2008. Double-blind test program for astrometric planet detection with Gaia, A\&A, 482, 699.

\bibitem{Forveille11} Forveille T., et al. 2011. The HARPS search for southern extra-solar planets XXXII. Only 4 planets in the Gl~581 system, A\&A submitted (arXiv:1109.2505).

\bibitem{Green85} Green R.M. 1985. Spherical Astronomy, Cambridge and New York, Cambridge University Press

\bibitem{Hatzes11} Hatzes A.P., et al. 2011. The Mass of CoRoT-7b, ApJ, 743, 75.

\bibitem{Lepine05} L\'epine S. 2005. Nearby Stars from the LSPM-North Proper-Motion Catalog. I. Main-Sequence Dwarfs and Giants within 33 Parsecs of the Sun, AJ, 130, 1680.

\bibitem{Marchal82} Marchal C., \& Bozis G. 1982. Hill Stability and Distance Curves for the General Three-Body Problem, Celest. Mech., 26, 311.

\bibitem{Sozzetti05} Sozzetti A. 2005. Astrometric Methods and Instrumentation to Identify and Characterize Extrasolar Planets: A Review, PASP, 117, 1021.

\bibitem{Sozzetti10} Sozzetti A. 2010. Detection and Characterization of Planetary Systems with $\mu$as Astrometry, EAS Pub. Ser., 42, 55.

\bibitem{Sozzetti11} Sozzetti A. 2011. Astrometry and Exoplanets: The Gaia Era and Beyond, EAS Pub. Ser., 45, 273.

\end{thebibliography}
\end{document}